# Photo Stabilization of p-i-n Perovskite Solar Cells with Bathocuproine: MXene

*Anastasia Yakusheva, Danila Saranin\*, Dmitry Muratov, Pavel Gostishchev, Hanna Pazniak, Alessia Di Vito,Thai Son Le, Lev Luchnikov, Anton Vasiliev, Dmitry Podgorny, Denis Kuznetsov, Sergey Didenko, Aldo Di Carlo\**

A.S. Yakusheva, D.S. Saranin, Pavel Gostischev, D.S. Muratov, T.S. Le, L.O. Luchnikov, A.A. Vasiliev, S.I. Didenko, Prof. A. Di Carlo

LASE – Laboratory of Advanced Solar Energy

National University of Science and Technology "MISiS,"

Leninsky prospect 4, 119049 Moscow, Russia

E-mail: saranin.ds@misis.ru

D.S. Muratov, D.V. Kuznetsov

Department of Functional Nanomaterials and High-Temperature Materials

National University of Science and Technology "MISiS,"

Leninsky prospect 4, 119049 Moscow, Russia

H. Pazniak

Université Grenoble Alpes, CNRS, Grenoble INP, LMGP

3 parvis Louis Néel, 38016, Grenoble, France

D.Podgorny

Department of Material Science in Semiconductors and Dielectrics,

National University of Science and Technology ''MISiS'',

Krymskiy val 3, 119049, Moscow, Russia

Alessia Di Vito, Prof. A. Di Carlo

C.H.O.S.E. (Centre for Hybrid and Organic Solar Energy), Department of Electronic Engineering

University of Rome Tor Vergata

via del Politecnico 1, 00133 Rome, Italy

E-mail: aldo.dicarlo@uniroma2.it

Keywords: MXenes, $Ti_3C_2T_x$, Inverted perovskite solar cells, Two-dimensional materials, Thermal stability, Photo stability

**Abstract**

Interface engineering is one of the promising strategies for the long-term stabilization of perovskite solar cells, preventing chemical decomposition induced by external agents and promoting fast charge transfer. Recently, MXenes – 2D structured transition metal carbides and nitrides with various functionalization (=O, -F, -OH) demonstrated high potential for mastering the work function in halide perovskite absorbers and significantly improved the n-type charge collection in solar cells. This work demonstrates that MXenes allow for efficient stabilization of perovskite solar cells besides improving their performances. We introduce a new mixed composite bathocuproine:MXene, i.e., (BCP:MXene) interlayer at the interface between an electron-transport layer (ETL) and a metal cathode in the p-i-n device structure. Our investigation demonstrates that the use of BCP:MXene interlayer slightly increases the power conversation efficiency (PCE) for PSCs (from 16.5 for reference to 17.5%) but dramatically improves the out of Glove-Box stability. Under ISOS-L-2 light soaking stress at 63± 1.5°C, The T80 (time needed to reduce efficiency down to 80% of the initial one) period increased from 460 h to > 2300 h.

1. Introduction

Halide perovskite solar cells (PSC) are one of the most promising photovoltaic technologies (PV).[1] Power conversion efficiency (PCE) of PSCs showed unprecedented growth during the last decade with record values of >25%.[2] Outstanding semiconductor properties, i.e., strong optical absorption in the visible part of the light spectrum ($>10^5$ $cm^{-1}$),[3] high diffusion length,[4] and mobility of the charge carriers,[5] could be achieved in microcrystalline thin films fabricated with low-cost solution-based methods. Among the different configurations, p-i-n architectures were considered of great importance due to the reduced impact of hysteresis in photoelectrical characteristics[6], [7] and the possibility of the low- temperatures (<200 ˚C) fabrication process.[8]

Generally, PSCs are heterostructure devices consisting of perovskite absorbers, charge-transporting layers (CTL), and electrodes. The quality of the heterointerfaces plays a critical role in the energy alignment between the functional layers, the efficiency of the charge collection, and the concentration of the defects. Interface engineering to reduce potential barriers and trap passivation was considered an effective way to surpass the efficiency of PSCs.[9] The electron transport layers in p-i-n PSCs play a crucial role in selective charge collection and chemical stability at contact with a metal electrode. The presence of charged ionic defects (iodides, ions of organic cations - $CH_3NH_3^+$, $CH_3(NH_2)^{2+}$, etc.) in perovskite absorbers causes their migration towards the metal cathode. It initiates the corrosion processes, typically accelerated by light and heat.

Fullerenes, such as $C_{60}$ or its soluble derivative [6,6]- phenyl-$C_{61}$-butyric acid methyl (PCBM), are the most widely used electron transport materials for the p-i-n structures of solar cells. This type of organic semiconductors was considered an effective passivation agent for the grain boundaries of microcrystalline halide perovskite films.[10], [11] Intrinsic $C_{60}$ based fullerenes demonstrates low values of the charge carrier mobilities ($10^{-2}$ - $10^{0}$ cm$^2$V$^{-1}$s$^{-1}$) and deep position of the lowest uncopied molecular orbit (LUMO) ~ - 4.5 eV[12] - 4.2 eV.[13] This reduces the charge collection efficiency at the interfaces with hybrid perovskites (such as MAPbI$_3$ and FAPbI$_3$ with minimum conduction band positions at -3.9 – 4.0 eV). Moreover, the hole blocking properties and the energy level alignment of fullerenes with metal cathodes are not ideal. Doping of C$_{60}$ films with the introduction of small organic molecules to the surface allows for a Fermi level shifting, a local increase of the conductivity,[14] and enhances the charge extraction in thin-film solar cells.[15]–[17] However, such a strategy requires complex and expensive processes to synthesize dopants. C$_{60}$-based materials are not stable enough to mitigate ion defects from a perovskite absorber to a metal electrode.[18] For this reason, some efforts have been made to develop cathode buffer films (CBFs) on the top of the ETL to improve charge collection, efficiency, and overall interface stability between ETL and back electrodes. The interface engineering strategies for the development of CBF include the use of polyelectrolytes (PEIE, PFN,[19], [20] etc.), metal multilayers,[21] and inorganic materials (typically oxides[22]). However, up to date, all the developed strategies for the implementation of new materials for the passivation of interfaces have not provided full stabilization of PSCs.[23]

Recently, a new perspective class of two-dimensional (2D) materials – MXenes provided a new effective route for interface engineering in perovskite solar cells PSCs.[24] In general, the MXenes are transition metal carbides/nitrides with single or multilayer 2D structures, with a general formula of M$_{n+1}$X$_n$T$_x$ (n = 1, 2, 3), where M represents an early transition metal, X is a carbon and/or nitrogen and T$_x$ are surface functionalization groups (–F, =O, –OH), which are inherited from the etching environment.[25] Such functionalization significantly influences the electronic structure of MXenes and provides a possibility for a work function tuning in the wide range of values (from 1.6 eV to 6.25 eV).[26] Moreover, the high intrinsic conductivity of MXenes provides excellent charge transport properties and opens possibilities for their application in optoelectronic devices.[27]–[29] The pioneer works [24], [30] on MXenes in n-i-p PSCs demonstrated enhanced electron transport efficiency induced by optimizing energy level alignment between a perovskite absorber and CTLs. Such an approach has been extended to p-i-n PSCs through doping a MAPbI$_3$ perovskite absorber and PCBM with MXenes, resulting in a boost of the output performance.[31] Until now, the impact of MXenes in PSCs was studied with respect to the charge transport improvement even though the rich surface chemistry of MXenes could provide passivation of the charged defects and lead to enhanced device stability. This paper investigates the performance stabilization of p-i-n PSCs due to Ti$_3$C$_2$T$_x$ MXene addition and discusses a possible mechanism to underline this effect. We developed a hybrid Bathocuproine(BCP):MXene interlayer, and it used as a composite film for electron charge transport and surface trap passivation at the interface with the metal cathode in p-i-n PSC with the following architecture: ITO/NiO$_x$/CsFAPbI$_3$/C$_{60}$/BCP:MXene/Cu. Our investigation demonstrated that the use of CBF based on BCP: MXene results in slight improvement of the PCE for PSCs (from 16.5 for reference to 17.5% for BCP:MXene device) but ~~dramatically~~ improves light soaking stabilities reaching T80>2000 h for the stress test.

## 2. Results and Discussions

### 2.1 $Ti_3C_2T_x$ MXene; BCP:MXene characterization and investigation of the surface morphology

We analyzed the morphology and electron diffraction patterns of $Ti_3C_2T_x$ MXene with transmission electron microscopy (see **Figure S1** of S.I.). Dried on the copper grid, MXenes form agglomerates with a flake-like structure and average size varying from 400 nm to 1 µm with different thicknesses, as visible by the change in the TEM image contrast. SAED patterns of $Ti_3C_2T_x$ show the hexagonal structure inherited from the parent MAX phase precursor. (**Figure S1a1, S1b1**).[32]

The surface chemistry of as-prepared $Ti_3C_2T_x$ MXenes was studied by X-ray photoelectron spectroscopy (XPS) (**Figure S2** and **Table S1**). The survey spectrum (**Figure S2a**) indicates the presence of the main elements of Ti, C, O, and F, and the traces of Cl in the surface composition of the MXenes, and no presence of Al, which confirms successful chemical etching. High-resolution (HR) XPS spectra of the Ti2p core levels of pristine $Ti_3C_2T_x$ (**Figure S2b**) were fitted with three doublets (Ti2p$_{3/2}$ and Ti2p$_{1/2}$) with the center peak at 454.7 eV attributed to Ti–C bonds.[33] The peak at higher energy (466.0 eV) belongs to C–Ti$^{+2/+3}$ and attributes to the surface functional groups. The presence of a peak at 458.4 eV corresponds to TiO$_2$ within the MXenes structure due to spontaneous air oxidation. The O1s core level shows peaks at 529.6, 530.76, and 532.08 eV, representing TiO$_2$, C–Ti–O$_x$, and C–Ti–OH$_x$, respectively (**Figure S2c**).[33] The C1s core level is fitted by three components centered at 281.6, 284.55, and 288.1 eV, corresponding to Ti–C, C–C, and COO, respectively (**Figure S2d**).[33]

Survey and corresponding HR spectra of BCP molecule mixed with $Ti_3C_2T_x$ MXenes (**Figure S2e-h**) show all the characteristic peaks of $Ti_3C_2T_x$, confirming its homogeneous distribution in BCP. Interestingly, the intensity of the C-Ti-(OH)$_x$ peak in HR O1s spectra (**Figure S2g**) is significantly increased as compared to pristine MXenes (**Figure S2c**). This could be assigned to the formation of hydrogen bonds by the interaction of BCP molecules with –O and –OH surface functional groups of MXenes. Moreover, the C1s peak (**Figure S2h**) is much broader when BCP is mixed with $Ti_3C_2T_x$ due to the possible modification in the energy levels of BCP molecule as a result of the above-mentioned interaction of BCP with abundant MXenes surface groups. After deposition of a thin $Ti_3C_2T_x$-BCP layer on top of C60 (**Figure S2i-l**), the C1s peak becomes dominated in the survey spectrum. At the same time, HR XPS spectra of Ti2p, O1s, and C1s resemble the one of BCP mixed with MXenes, suggesting their stability while combining with the C60 layer. In our work we used $Ti_3C_2T_x$ MXenes produced under MILD etching conditions. The valence band XPS (**Fig. S2n**) of synthesized $Ti_3C_2T_x$ MXenes show that the valence band maximum is at the Fermi level. This confirm the metallic character of MXenes originating from Ti3d states which are at E$_f$. The interaction of $Ti_3C_2T_x$ with BCP molecules additionally could be confirmed by the XRD patterns shown in **Figure 1**. Adding a small amount of MXenes (10 vol.%) to the BCP results in the presence of a visible characteristic (002) low angle MXene peak, which is shifted to low 2θ values and exhibits a broader shape in comparison with as-prepared $Ti_3C_2T_x$.

This is the evidence of an expansion of the interlayer distance between the $Ti_3C_2T_x$ layers (from 13.49 to 14.45 Å) resulting from the intercalation of organic BCP molecules and the formation of the $Ti_3C_2T_x$-BCP heterostructure. This is possible since the fully terminated negatively charged $Ti_3C_2T_x$ surface easily interacts with the polar functional groups of BCP molecules, ensuring a uniform distribution of Ti₃C₂Tₓ in the organic and resulting in a homogeneous composite layer (**Figure S3**). To confirm the BCP molecules' impact on the interlayer distance between the MXene layers, we performed Density-functional theory (DFT) calculations on pristine MXenes and MXenes with the BCP molecules intercalated between the layers. The agreement of the DFT calculation for the experimental data is rather impressive: the calculated inter-distance of the pristine MXene layers is 13.465 Å, while it increases to 14.396 Å when the BCP molecule is intercalated between the layers. We can safely assume that the structural properties obtained by DFT calculations are consistent with experimental observations with an uncertainty that is approximately 1% (see **table S1 and additional details in the supplementary** information – S.I.). Note that with respect to the previous DFT studies,[34] where the BCP/metal interaction is treated assuming a parallel flat-lying molecule configuration on top of the metal surface, the situation here is quite different. In our case, due to the fully terminated surface of the Ti₃C₂Tₓ, the BCP molecules interact with the functional groups of MXenes, in particular F. The interaction of such coupled system is quite strong and induces a tilting of the BCP phenyl rings, yielding a non-completely flat-lying final configuration of the molecule. The DFT calculations considering MXenes with oxygen or OH as termination groups (**Figure S4a-d**) show that the final inter-distance between the MXenes layers with BCP does not vary significantly with respect to -F termination. On the other hand, the interlayer distance between the pristine MXenes layers reduces with -OH and =O terminations. In addition, the final configuration of the BCP molecule is completely flat-lying for -OH terminations, while the rotation of the phenyl rings is slight for =O and -F terminations (**Figure S4a-c**).

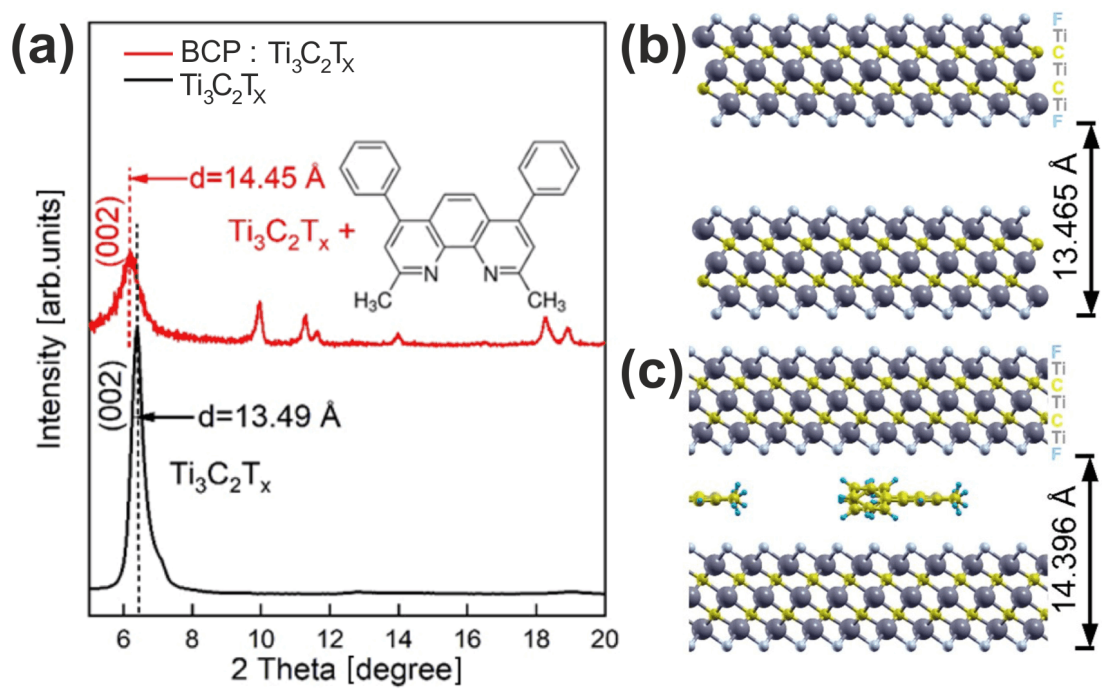

**Figure 1.** (a) XRD patterns of the Ti₃C₂Tₓ-BCP heterostructure in comparison with the Ti₃C₂Tₓ layers; the DFT simulations for the interlayer distance in (b) Ti₃C₂F and (c) BCP: Ti₃C₂F

## 2.2 Performance of PSCs with BCP:MXene interlayer

After the initial assessment of the surface and structural properties of MXenes and their composites with BCP, we performed a thorough analysis of the output IV performance and the long-term stability of PSCs. The initial hypothesis and the motivation for using MXenes at the interface with the metal electrode were to reduce chemical interaction between the electrode and the mobile ionic defects formed in perovskite films and diffusing towards the cathode through ETL. The inverted planar architecture (p-i-n orientation see **Figure 2a**) was realized using the $Cs_{0.2}FA_{0.8}PbI_x$ perovskite as an absorber layer (FA=Formamidinium), $NiO_x$ as HTL, C60 as ETL, MXene – BCP composite as a hole blocking interlayer, and a copper cathode. The schematic energetic diagram of the device is shown in **Figure 2b**, where the values for the energy levels were taken from the literature. [32]–[35]

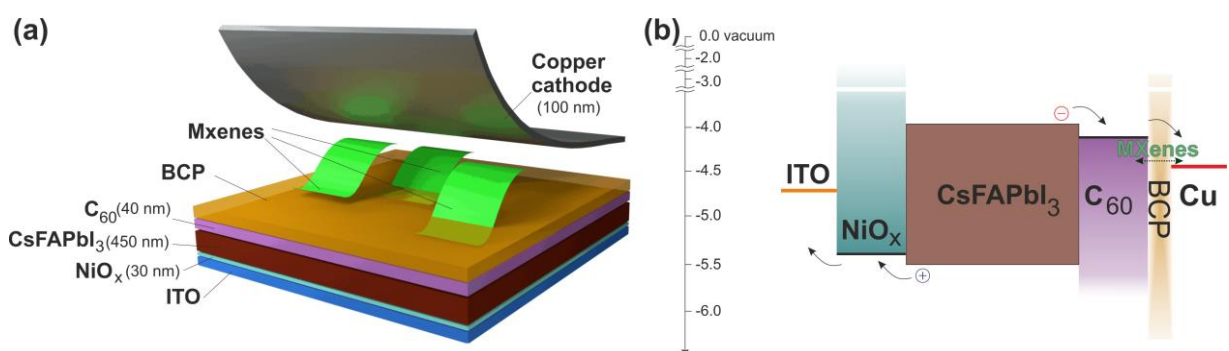

**Figure 2.** Device schematics with the MXenes-BCP interlayer (a), schematic energetic diagram of the studied PSCs (b)

First, we investigated the relation between the concentration of MXenes in BCP and the photovoltaic performance of PSCs to identify the optimized fabrication parameters. The precursor for the composite interlayer was obtained via ultrasonic dispersing of the MXene powder in BCP ink in IPA. The PSCs fabricated with pristine BCP were used as references. For the BCP:MXene device, we used several concentrations of MXenes in a range of 0.5 to 5.0 mg/ml (dispersion) in BCP ink (0.5 mg/ml in IPA, respectively). In the following, we will refer to the PSCs fabricated with different concentrations of MXenes as MX (0.50) for 0.50 mg/ml, MX (0.75) for 0.75 mg/ml, etc. To estimate the possible changes in thickness of composite CBF films we used stylus profilometry. The statistical analysis of the data showed that average thickness of CBF slightly increase by adding MXenes: from 9.1nm for MX (0.50) till 10.9nm for MX (5.00) (**See table S2 in S.I.).** The resulting thicknesses of the CBF composite films were comparable to the reference BCP layer (8.6nm) and varied in the published ranges of BCP thicknesses (8-11 nm)[36], [37]. Moreover, Atomic Force Microscopy (AFM, See **Figure S5 in S.I.**) analyzes show that the MX (0.75) CBF provides more uniform coverage of the $C_{60}$ surface.

The measurement of the PL quenching is one of the routine methods for the analysis of photo-injection properties in the multilayer thin-film stacks[38], [39], however, interpreting such effects is not always straightforward[40], [41]. We measured the PL spectrums of the multilayer stack (glass/perovskite/$C_{60}$/CBF) by varying the concentration of MXenes into BCP (**fig. S6** in S.I.). ~~Using CBF on the top of $C_{60}$, the PL is slightly quenched, that may be due to an improved photoinjecting or transport properties. Interestingly, the strongest PL quenching occurs for the multilayer stack with MX(0.75), while for the samples fabricated with an other concentration of MXenes the quenching was weaker.~~ Using CBF on the top of $C_{60}$, the PL is slightly quenched. Interestingly, the strongest PL quenching occurs for the multilayer stack with MX(0.75), while for the samples fabricated with other concentrations of MXenes the quenching is weaker. As pointed out by Campanari et al.,[42] it is not possible to make a general direct correlation between cell performances and quenching of the PL, however quenching may occur due to an improved photoinjecting or transport properties.

The photovoltaic (PV) parameters for the fabricated PSCs (open-circuit voltage - $V_{oc}$, short circuit current density - $J_{sc}$, fill factor–FF, and power conversation efficiency—PCE) were calculated and statistically analyzed from the IV characteristics measured under standard conditions of 1.5 AM G light spectrum with 100 mW/cm$^2$ irradiance power density (presented in the box charts in **Figure 3**). The reference devices demonstrated the average PCE values of (15.1±0.66)% with $V_{oc}$ =(1.01±0.01) V, $J_{sc}$ =(20.60±0.87) mA/cm$^2$, and FF = (69±3)%. The improvement in the output performance of the PSCs with the BCP:MXene interlayer was achieved only for small concentrations of MXenes (≤ 0.75 mg/ml). We found that the device configuration MX (0.75) demonstrated the best IV performance among the other investigated device configurations. The average output parameters of the MX (0.75) based PSCs demonstrated an increase of all output parameters in comparison with the reference cells with $V_{oc}$= (1.04±0.01) V with a +0.03 V with respect to the reference cell, $J_{SC}$ = (22.22±0.65) mA/cm$^2$ (+1.05 mA/cm$^2$), FF= (77±3)% (+0.8%), and PCE=(16.2±0.63)% (+0.9%). The IV curves for the best performing PSC with MX (0.75) are reported in **Figure 4a**. The champion reference device showed a PCE=16.45%, while the best BCP:MXene (0.75 mg/ml) PSC achieved a PCE=17.46%. The further increase of the MXene additive concentration >0.75 mg/ml in the BCP ink resulted in a sharp drop in the output performance mainly related to JSC losses and with a minor extend to $V_{oc}$. All device configurations demonstrated very negligible hysteresis (see **Fig.S7**). Figure 3 shows a strong impact of MXenes concentration on both $J_{sc}$ and FF. By analyzing J-V characteristics (reported in Figure **S8** (S.I.)) we observe that by increasing the MXene concentration till MX (5.00) a strong shunt effect appears. In fact, there is a clear relation between the series and shunt resistances dependence on the MXenes concentrations (see Figure **S9** in S.I.) with the $J_{sc}$ and FF dependence on MXenes concentration of Figure 3. The series resistance has a minimum for MX (0.75) while the shunt resistance has a maximum for MX (1.00). The reduction of shunt resistance is most probably related to the effect of clustering of the metallic MXenes at high concentration as shown by the AFM images of Figure S5. On the other hand, such clustering reduces the MXenes dispersion into the BCP by increasing series resistance (see Fig. **S10**, table **S3** in **S.I**. and discussion below on CBF resistivity). To estimate the hole blocking properties of MXenes flakes, we fabricated and characterized PSCs with pure MXene interlayer without BCP. For this configuration, the PCE values of all fabricated devices are below 10%. Mostly, the drop of the output performance was caused by losses in FF and $V_{oc}$ (see **the fig.S11 in S.I.**) related to an increased impact of the non-radiative recombination at the cathode interface and to resistive shunts. AFM analysis proved that MXene flakes and aggregates do not from continuous thin-film, thus we assume that the use of BCP for composite CBF is necessary to provide isolation of metal/$C_{60}$ junction. Therefore, we considered that the use of pure MXene interlayer is irrelevant and excludes this device configuration from the further experiment.

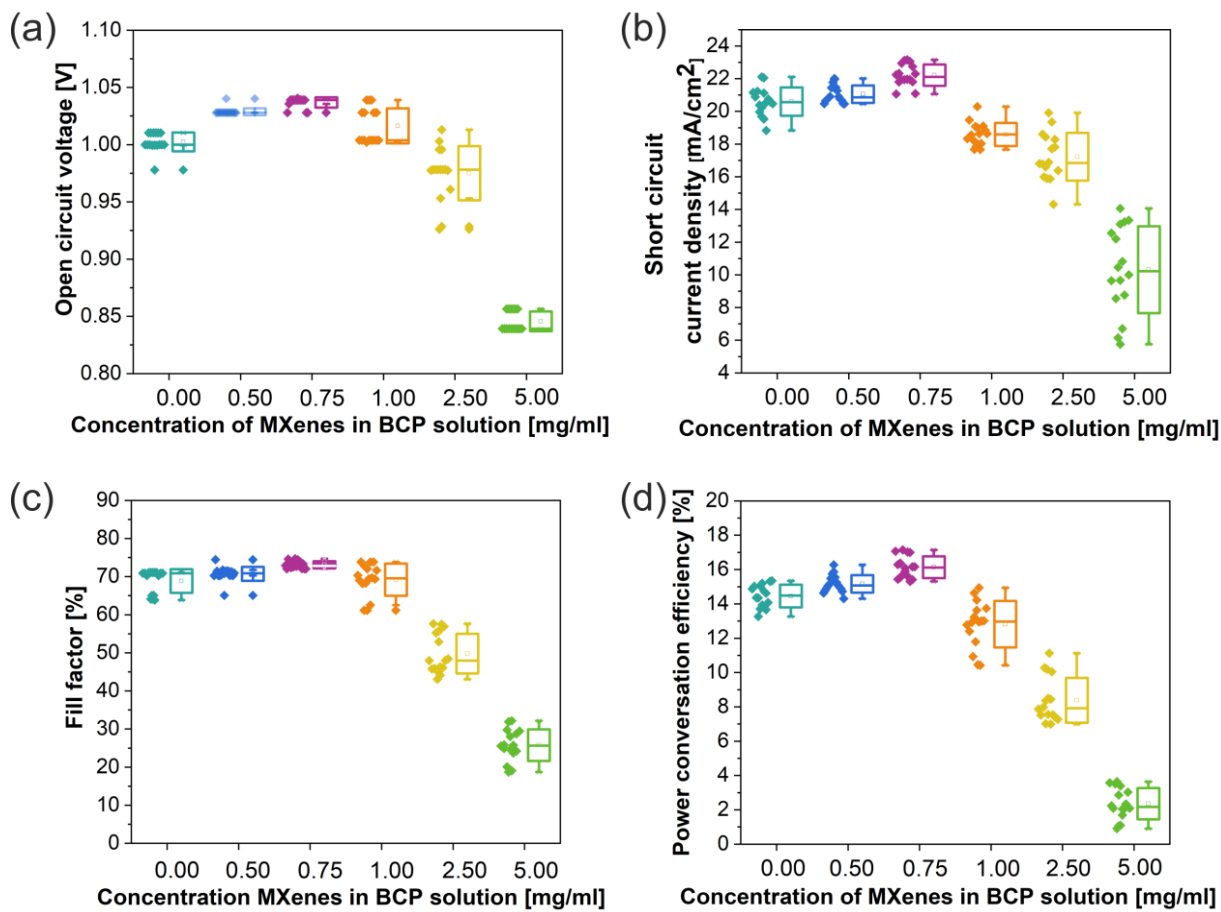

**Figure 3.** Box charts for the IV output parameters of the PCSs fabricated with different concentrations of MXenes in BCP ink – Open circuit voltage (a), short current density (b), fill factor (c), power conversation efficiency (d)

Generally, the $V_{oc}$ is related to splitting the quasi-Fermi levels (QFL) in the absorber and the charge-transporting layers, resulting in a careful balance between photon absorption and charge recombination. In this work, we improved the $V_{oc}$ values (+4% concerning reference devices) for MX (0.75). As shown with the device calculations in the previous works,[31] the main role of the MXenes at the ETL side is the reduction of interface recombination and the alignment of energy levels with a metal electrode. The external quantum efficiency measurements (EQE, **Figure 4b**) show an overall increase in all spectral ranges. We relate this improvement of the EQE with the enhanced efficiency of electron collection. Improvement of charge extraction can also be related to the buffer layer's variation of conductivity. To estimate the impact of the MXenes doping of BCP on the resistivity of the layer, we measured the IV curves (**Figure S10**) for a resistor structure ITO/ETL/CBF/Cu where CBF is BCP or BCP:MXene. The specific resistivity ($\rho$) of the reference C60/BCP stacks between ITO and Cu electrodes has an average value of $1.9 \times 10^5$ ($\pm 0.06 \times 10^5$) Ohm*cm$^{-1}$, while the presence of MXenes significantly reduces the resistivity of the C$_{60}$/BCP:MXene layer down to $\rho =6.0 \times 10^4$ ($\pm 0.25 \times 10^4$) for MX(0.75), which is more than three times smaller with respect to the reference device (**Table S3**). Using CBF with high concentration of MXenes (MX (2.50) and MX (5.00)) increased the values of the specific resistivity up to $7.21 \times 10^4$ ($\pm 0.47 \times 10^4$) and $2.59 \times 10^5$ ($\pm 0.67 \times 10^4$) Ohm*cm$^{-1}$ respectively due to the mentioned clustering effect.

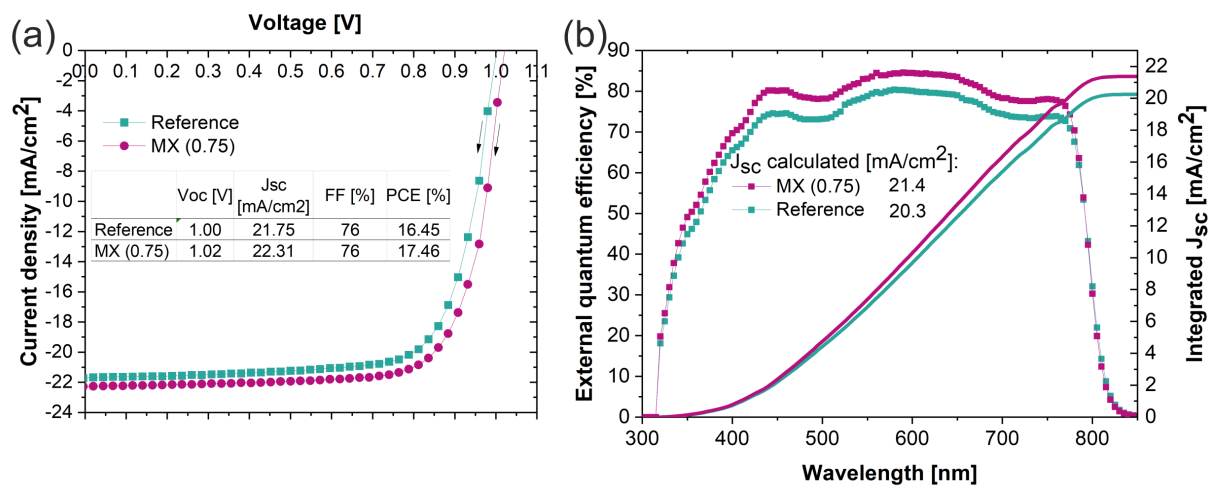

**Figure 4.** IV curves for the best performing PCSs (a), external quantum efficiencies spectrums for the references and PSCs with BCP:MXene

Besides improving the output performance of PSCs, the main benefit for the use of BCP:MXene composite is manifested in the improved stability of the device operation under the photo- and thermal- stress with respect to reference devices. The stability test of PSCs was performed in air with glass-glass encapsulated devices (see experimental for more details) following the ISOS protocol. [43] To compare the stability of different devices, we consider the $T_{80}$ parameter, which is the time to reach 80% of the initial performance. Thermal stability of the PCSs was studied according to the **ISOS-T-1** with storage of the devices in the dark oven heated at 80 ˚C (open circuit conditions) and periodical IV characterization (standard 1.5 AM G conditions).

The thermal stability of the PSCs' PV parameters is well correlated to the concentration of MXenes used for the composite interlayer. Using a higher concentration of MXenes for the fabrication of CBF resulted in higher thermal stability and longer T80 time, as shown in **Figure 5a** for the device PCE (the $V_{oc}$, Jsc, FF dependence on the MXenes concentration is presented in **Figure S12**).

The stabilization effect of MXenes permits increasing $T_{80}$ of the PCE from 330 h of the reference cell to 1080 h for the BCP:MXene CBL with MX (1.0), while for the MX (0.75) device, T80 is 790 h. It demonstrates straight dependence on the reduction of the degradation dynamics to the concentration of MXenes in CBF under thermal stress. The changes in the PCE were mainly related to the decrease of $J_{sc}$ and FF, while the $V_{oc}$ was quite stable for all the device configurations (**Figure S12a**) with a reduction of around 4-8% after 1300 h of stress. The reference devices reached $T_{80}$ for $J_{sc}$ of 1100 h, while only MX (1.0) showed a clear improvement of such value with a $J_{sc}$ decrement of 14% after 1300 h of thermal stress (**Figure S12b**). By contrast, the increasing concentration of MXenes into BCP on the FF stability is clearly evident (**Figure S12c**). The reference PSCs have $T_{80}$ for FF equal to 760 h, while for the MX (0.50) and MX (0.75) devices, this value increased up to 1050 and 1200 h, respectively, and it was beyond 1300 h for MX (1.0). As shown by AFM images (Fig. S5 in S.I.), MX (1.00) provides more dense coverage of the ETL surface with respect to the MX (0.75). The maximum aggregate size of MXene flakes for MX (0.75) is 135 nm, while for MX (1.00) this value increases to 221 nm. As discussed above, the agglomeration of MXenes could impact on the cell performance through changes in the shunt and series resistances. Despite this, better coverage of the ETL surface with BCP:MXenes could provide higher protection efficiency against ion diffusion process enhanced by the heat. Thus, further improvement of the performance and thermal stability of the PSCs using MXenes at the interfaces, requires optimization of the deposition methods providing uniform coating of the 2D flakes without aggregates and pinholes.

Light Soaking stability (ISOS-L-2) was assessed considering the MPPT of PSCs under LED illumination (**Figure 5b,** see experimental for the MPPT algorithm) for the most efficient device configuration MX (0.75) and compared to the reference PSCs. The continuous application of positive bias (Vmax), light and heat stress (the temperature of the tested PSCs was 63±1.5 ˚C) resulted in a quite fast reduction of Pmax values of the reference PSCs with T80 of 430 h. The use of the BCP:MXene interlayer ~~enormously~~ substantially enhanced the light soaking stability with an average reduction of the output power with respect to the initial one of only 4% after the 2300 h of the stress test. Thus, T80 should vastly exceed 2300 h (at least five times larger than the reference).

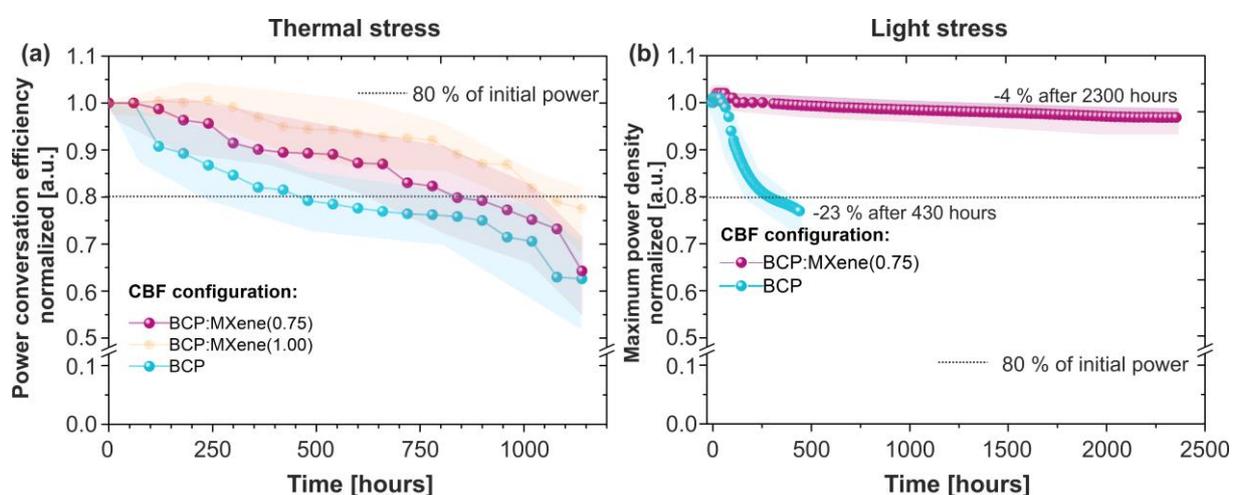

**Figure 5.** Dependence of the PCE to the time of (a) thermal stress (ISOS-T-1 for the encapsulated devices) with respect to the used MX concentration (b) MPPT for the best performing configuration of the MX doped PSCs – MX (0.75) in comparison with the reference device configuration (ISOS-L-2). The lines and the symbols represent the performance of the most stable devices; the semitransparent cloud areas represent the standard deviation of the parameters for the tested PSCs

To clarify the role of MXenes addition to BCP on the stability of the PSC, we measure the dark IV characteristics (**Figures S13 and S14**) of reference and MX (0.75) PSCs during thermal and light soaking stress. The dark IV characteristics of the two PSCs evolve under stress in a very different manner. In particular, the increase of the dark saturation current ($J_0$) during the stress is reduced in MX (0.75) with respect to reference devices. Considering that $J_0$ is ruled by charge recombination,[40] we can assert that MXene addition to BCP prevents the formation of new recombination centers into the perovskite layer during the stress test as it occurs for the device without MXene. The chemical activity at the interfaces was considered one of the significant factors for the degradation of PSC and its output performance. [9], [44] The migration of under-coordinated and volatile bimolecular iodine ($I^-$); ($I_2$)[45], [46] can cause irreversible decomposition of perovskite absorber with a production of lead oxides, hydroxides, and $PbI_2$. This structural and interfacial degradation of the halide perovskite thin-films is recognized as typical processes occurring under continuous light absorption and heating conditions. [47], [48]

The high density of the ionic defects with low activation energy diffuses and initiates the electrochemical corrosion of charge-transporting layers[49], electrodes[50], and their interfaces. The $C_{60}$ based n-type organic semiconductors provide excellent charge transport in PSCs and passivate the surface defects of perovskite absorbers but do not provide a diffusion barrier of iodine-containing defects,[51], [52] which migrate across the thin film up to the metal electrode.[53]

An additional element deserves some attention, namely the increase of BCP conductivity when MXenes are added. *Ahn and co-workers* described [54] the mechanism of trapped charge induced degradation of perovskite absorbers at the interfaces with CTLs. The charge extraction rate of CTL could impact the trapping and decomposition processes. Slow extraction of the holes/electrons tends to accumulate charges at the perovskite/CTL interfaces, while faster extraction mitigates this process. At the same time, the unbalanced charge extraction at the hole and electron transport interfaces could induce the charge accumulation at the interface with the CTL with a low value of charge carrier mobility[54], [55]. The results of improved stability in PSCs with high-mobility inorganic CTLs support this argument.[56] In our case, the use of BCP:MXene for p-i-n PSCs significantly increases the conductivity of the organic ETL stack ($C_{60}$/BCP), reducing the charge accumulation at the perovskite/ETL interface and consequently the device degradation. To further elucidate the impact of CBF on the fabricated PSCs we performed electrochemical impedance spectroscopy (EIS). Varying from BCP to MXene:BCP based PSCs, the Recombination resistance ($R_{rec}$) increases (see **Fig. S15 in S.I.**). When MXenes are added to BCP, the increased conductivity of the CBF decreased the impact of the charge carrier trapping at the ETL interface.

The chemical activity at the interfaces was considered as one of the major factors for the degradation of PSC and its output performance[57]. The migration of under-coordinated and volatile bimolecular iodine ($I^-$); ($I_2$)[58], [59] can cause irreversible decomposition of perovskite absorber with production of lead oxides, hydroxides[60] and $PbI_2$[61]. This structural and interfacial degradation of the halide perovskite thin-films is recognized as typical processes occurring under conditions of continuous light absorption[62] and heating[63]. The high density of the ionic defects with low activation energy diffuse and initiate the electrochemical corrosion of charge transporting layers[64], electrodes[65] and their interfaces. $C_{60}$ based n-type organic semiconductors provide excellent charge transport in PSCs and passivate the surface defects of perovskite absorber[66], but does not provide diffusion barrier of iodine containing defects[67], that migrate across the thin film up to metal electrode[68]. To investigate the impact of composite hole blocking interlayers BCP:MXene on the ion diffusion we performed Auger-elemental profiling for model devices with and without MXenes and for both as-fabricated (fresh) and thermally stressed structures (Fig. **S16 in S.I.**). Auger profiles shows that both reference (BCP) and BCP:MXene fresh samples have very similar profiles for all analyzed elements (Ag, Pb, I, Si, see Figs. S16a and S16b). On the other hands, thermally stressed samples show significant widening of iodine profiles on the reference structures, while the BCP:Mxene structure does not (Figs. S16c, S16d). Thus, the Auger analyses suggests that MXenes:BCP composite decrease the rate of degradation of the interface limiting iodine migration.

3. Conclusion

This work demonstrated a new strategy for long-term stabilization of p-i-n PSCs with integrating a bathocuproine:MXene composite interlayer at the metal cathode interface. A special cathode buffer layer was realized with solution processing dispersing of MXenes into a BCP solution at different concentrations. We found that the role of BCP in composite with MXenes is essential for the uniform distribution of $Ti_3C_2T_x$ in the formation of a homogeneous surface layer. The use of MXenes with a concentration of 0.75 mg/ml was found to be optimal to (slightly) improve the output performance for $CsFAPbI_3$ – based PSCs with a PCE increase from 16.45 % of the reference PSC to 17.46% for the BCP:MXene device. This improvement is mainly due to the higher efficiency of the electron collection and the improved conductivity of the ETL stack. However, the main contribution of the BCP:MXene interlayer was related to the stabilization of the device. The thermal stability test at 80 °C (tested with ISOS-T-1 procedure) showed a remarkable increase of T80 of the BCP:MXene device up to 1080 h with respect to the reference PSC (T80 = 330 h). This was even more emphasized in the light soaking stability at MPPT in the air (encapsulated devices), where the BCP:MXene perovskite solar cells showed a reduction of only 4% of the output power after 2300 h of continuous light soaking, whereas the reference device showed only $T_{80}$ of 430 h.

The analysis of dark IV curves showed that BCP:MXene composite stabilizes shunt properties and rectification in the device under stress factors. The origin of the stabilization using BCP:MXene composite occurs in rich chemistry of interaction between low-dimensional materials, small organic molecules, and ionic defects formed in perovskite. The BCP:MXene interaction with $C_{60}$ and the perovskite layers results in passivation of the surface states of MXenes, which could inhibit the degradation of the whole structure in a long-term perspective. We assume that advanced surface termination between MXenes- BCP, and $C_{60}$ suppresses the chemical interaction with mobile ionic defects at the cathode side of the PSCs. In addition, the enhancement of the ETL stack conductivity when BCP:MXene is used can reduce the trapped charges at the perovskite/ETL interface and consequently decrease the impact of trapped charge-driven degradation. As a final element, Auger profile analyses show, for thermally stressed device, a reduction of iodine migration for MXene:BCP with respect to pristine BCP.

## 4. Experimental Section

### 4.1 MXenes preparation

The $Ti_3C_2T_x$ MXenes were synthesized by selective chemical etching of Al layers from a powder $Ti_3AlC_2$ MAX phase precursor (particle size < 38 µm) using a minimally intensive layer delamination (MILD) approach. Briefly, 0.5 g of the $Ti_3AlC_2$ MAX phase was continuously added in a 10 mL solution of 0.8 g of LiF dissolved in 9 M HCl under constant magnetic stirring during 24 h at RT. The obtained suspension was centrifuged at least five times at 3500 rpm for 5 min until neutral pH (6-7) of the solution was reached. The obtained solution of well-delaminated $Ti_3C_2T_x$ MXenes was vacuum-filtered and dried at 80 ºC at a vacuum atmosphere for 24 h.

### 4.2 MXene/BCP layer composite

The BCP solution was prepared in 0.5 mg/ml concentration in anhydrous isopropyl alcohol (IPA), whereas MXenes were taken in five different weight concentrations: 0.5, 0.75, 1, 2.5, and 5 mg per ml of the BCP solution. The mixture was kept for 1-2 days stirred by a magnetic stirring bar on a hotplate at 50 ˚C for preliminary BCP coordination around the MXene surface. The BCP interlayer was spin-coated at 4000 RPMs (30 s) and annealed at 50 °C (1 min).

### 4.3 Materials for ink preparation

Herein, the organic solvents dimethylformamide (DMF), N-Methyl-2-pyrrolidone (NMP), chlorobenzene (CB), 2-propanol (IPA), and methoxyethanol (MOE) were used in anhydrous, ultra-pure grade from Sigma Aldrich (Germany). The device was fabricated on $In_2O_3$:$SnO_2$ (ITO, 10 ohm/sq) coated glass from Kaivo. We used acidified nickel chloride ($NiCl_2$) as a precursor for the NiO hole-transporting layer (HTL). Lead Iodide (PbI2 99.9 % purity LLC Chemosynthesis (Russia)) and cesium iodide (CsI 99,9 %, Lanhit (Russia)) formamidinium iodide (FAI, 99.99 % purity from GreatcellSolar) were used to prepare the perovskite ink. Fullerene (C60 99.9 % MST, Latvia) and bathocuproine (BCP, >99.5 % sublimed grade, Osilla Inc., UK) were used as an electron-transporting layer (ETL).

### 4.4 Materials for PSC preparation

The p-i-n-structured solar cell with the following stack ITO/NiOx/CsFAPbI3/C60/BCP/Cu was fabricated via an optimized route. Firstly, the ITO substrates were cleaned with detergent, de-ionized water, acetone, and IPA in an ultrasonic bath. Then the substrates were activated under UV-ozone irradiation for 30 min. The solution of nickel nitrates 0.15 M in MOE for NiO HTL was spin-coated at 4000 RPMs (30 s), dried at 120 °C (10 min), and annealed at 300 °C (1 h) in ambient atmosphere. The CsFAPbI3 film was crystallized on top of HTL with a solvent-engineering method. The perovskite precursor was spin-coated with the following ramp: (1s – 1000 rpm, 4 sec – 3000 rpm / 30 sec – 5500 rpm). In details, 420 µL of CB were dropped on the substrate on the 10th second after starting the first rotation step. Then the substrates were annealed at 70 °C (1 min) and 105°C (30 min) to form the appropriate perovskite phase. C60 was deposited with the thermal evaporation method at $10^{-6}$ Torr vacuum level. The free BCP interlayer was spin-coated at 4000 RPMs (30 s) and annealed at 50 °C (1 min) for the reference devices. The copper cathode was also deposited with thermal evaporation through a shadow mask to form a 0.14 $cm^2$ active area for each pixel. All fabricated devices were encapsulated with a UV curable epoxy from Osilla Inc. and a glass coverslip.

### 4.5 Characterization Methods of MXene and MXene/BCP layer

The X-ray diffraction patterns of the $Ti_3C_2T_x$ MXene and $Ti_3C_2T_x$-BCP heterostructures were recorded using a Malvern Panalytical X'Pert MPD PW3040 diffractometer (Cu anode, $\lambda$ = 0.15418 nm) in 10°-60° 2θ-range with steps of 0.002°. The microphotographs and electron diffraction patterns were obtained with a Transmission electron microscope JEOL JEM-1400 (Japan) for the morphology and crystal structure analysis of MXenes. The X-ray photoelectron spectroscopy (XPS) was performed with an electronic spectrometer Kratos AXIS Ultra DLD (Kratos Analytical Ltd, UK) for the surface chemical analysis of the termination group and the valence band position measurements in the $Ti_3C_2T_x$. Also, XPS was used to analyze the BCP/MXnene composite structure and the chemical interaction between the MXenes and C60 layers. Further characterization was provided by scanning electron microscopy (SEM) using JEOL JSM-6610LV with Oxford Instruments energy dispersive spectroscopy (EDS) attachment: AZTEC EDS Energy Max 80) for the BCP+MXene film in the MXene morphology in the device from the ITO side (ITO/NiO/Perovskite/C60/BCP+MX) and the elemental analysis for evaluating the conductive area due to the inclusion of MXene.

The system's atomic positions and lattice parameters were optimized using the Broyden-Fletcher-Goldfarb-Shanno (BFGS) quasi-newton algorithm implemented in the QuantumEspresso software.[49] We employed the scalar-relativistic, Perdew-Burke-Ernzerhof (PBE)[50], projector-augmented-wave (PAW) pseudo-potentials available in the QuantumEspresso library.[51] For both the MXene and the BCP/MXene systems, we set the kinetic energy cutoff of wavefunctions to 70Ry and the convergence threshold on forces to $10^{-2}$Ry/Bohr.

For the geometric optimization of the $Ti_3C_2F_2$ MXene, we used an 8×8×2 Monkhorst-Pack k-points grid and a convergence threshold on the total energy of $10^{-4}$Ry (default value). For the optimization of the BCP/MXene supercell, composed of 6×6 $Ti_3C_2F_2$ unit cells (252 atoms) and a BCP molecule (48 atoms) placed in the middle between two MXene layers, we considered only the Gamma point and a convergence threshold on the total energy of $10^{-3}$Ry (note that this is an extensive property, like the total energy, and the total number of atoms in the supercell is more than 40 times larger than the number of atoms in the MXene unit cell).

*4.6 The Device characterization methods*

The photovoltaic (PV) parameters for the fabricated PSCs (open-circuit voltage - Voc, short circuit current density - Jsc, fill factor – FF, and power conversation efficiency – PCE) were calculated and statistically analyzed from the IV characteristics measured under standard conditions of 1.5 AM G light spectrum. We used Xenon arc-lamp based solar simulator (ABET 3000) with AAA grade of conformity to the reference Air Mass 1.5 Global spectrum of the terrestrial solar light (1.5 AM G standard - (ASTM) G-173). The standard power density of the incoming light (100 mW/cm2) was calibrated with certified Si- reference solar cell. The scan sweeps and the MPPT tuning measurements were performed with Keithley 2400 SMU. The measurements of IV curves were done with use of the aperture mask (3x3 mm$^2$). For measurements of the external quantum efficiency, we used QEX10 Quantum Efficiency System based on the broadband Xe-light source monochromator and adjustable mechanical chopper. The conformity of spectral response for the measured PSCs was calibrated with Si-solar cell and was compliant to the ASTM E 1021-06 standard. The difference between Jsc values gathered from IV and those extracted from EQE measurements is mainly related to the different performance of the certified calibration cells used for adjustments to the standard conditions of illumination for solar simulator and EQE system. The stability tests were performed with an LED light source (4000 K) equilibrated to the 1 sun conditions with alignment to $J_{sc}$ values. The MPPT tests were done using an automated measurement system controlled by LABVIEW software. The algorithm for MPPT included the initial IV scan in the forward direction (from 0 V up to $V_{oc}$ value, sweep step of 10 mV), calculation of maximum power, tracking of the changes for $V_{max}$, $I_{max}$, and $P_{max}$ with time (1 update per second). The cycle for MPPT was performed for 60 min and then repeated. The thermal stability of the PCSs was studied according to the ISOS-T-1 with storage of the devices in the dark oven heated at 80 ˚C (open circuit conditions) and periodical IV characterization (every 24 - 48 h under standard 1.5 AM G conditions). To analyze the internal processes and elucidate the impact of CBF on the fabricated PSCs, we used electrochemical impedance spectroscopy (EIS). Measurements were performed using Keysight E4980A Precision LCR-meter applying 0 V constant bias with superimposed 0.1 V AC through at 20 Hz to 2 MHz frequency band.

## Acknowledgments

The authors gratefully acknowledge financial support from the Ministry of Science and Higher Education of the Russian Federation in the framework of Mega Grant [No. 075-15-2021-635].

# Conflict of Interest

The authors declare that they have no known competing financial interests or personal relationships that could have appeared to influence the work reported in this paper.

# Abbreviations

BCP:MXene – MXene and bathocuproine in composite

BCP – bathocuproine

CTL – charge-transporting layer

CBF – cathode buffer film

DFT – Density-functional theory

ETL – electron-transport layer

IPA – isopropyl alcohol

PCE – power conversation efficiency

PSC – perovskite solar cell

PV – photovoltaics

PCBM – [6,6]- phenyl-C61-butyric acid methyl

QFL – quasi-Fermi levels